\documentstyle[psfig]{article}

\begin{document}

\title{The continuum structure of the Borromean halo nucleus $^{11}$Li }

\author{A. Cobis, D. V. Fedorov and A. S. Jensen\\
Institute of Physics and Astronomy, Aarhus University,\\
DK-8000 Aarhus C, Denmark}

\date{}
\maketitle

\begin{abstract}
We solve the Faddeev equations for $^{11}$Li (n+n+$^9$Li) using
hyperspherical coordinates and analytical expressions for distances
much larger than the effective ranges of the interactions.  The lowest
resonances are found at 0.65 MeV
($\frac{1}{2}^{+},\frac{3}{2}^{+},\frac{5}{2}^{+}$) and 0.89 MeV
($\frac{3}{2}^{\pm}$) with widths of about 0.35 MeV. A number of
higher-lying broader resonances are also obtained and related to the
Efimov effect. The dipole strength function and the Coulomb
dissociation cross section are also calculated.
\end{abstract}

\paragraph*{Introduction.} The new class of nuclear states, called
halos, are presently intensively investigated \cite{han95}. Their
structure and behavior in reactions differ substantially from those of
ordinary nuclei. They are unusually large and loosely bound. Most of
them are presumably not identified yet. Large electromagnetic
dissociation cross section, or a concentration of $1^{-}$ strength
function at low energies exemplify the unusual character of these
nuclei. Descriptions as two- and three-body systems have been
successful \cite{fed95,gar96}. The continuum structure of halo nuclei
is still almost unknown even for the most studied halo nucleus
$^{11}$Li. However almost inevitably low-lying resonances must be
present, since the Borromean nature implies that the two-body
subsystems are close to threshold and perhaps nearly fulfill the Efimov
condition \cite{efi70,fed93}.

A recent experiment, where protons are scattered on $^{11}$Li, was
interpreted as evidence for a $1^{-}$-resonance in $^{11}$Li at 1.0
$\pm$ 0.1 MeV above the two-neutron threshold with a width of about
0.75 $\pm 0.6$ MeV \cite {kor97}. An other interpretation of the same
experiment is given in terms of the shake-off process
\cite{kar97}. Reaction experiments also indicate an excited
$^{11}$Li-state at about the same energy \cite{kor97,zin97}.  The
measured \cite{zin97,sac93,shi95} and computed
\cite{esb92,dan94,sag96} $1^{-}$-strength functions are available in
the literature, where the absence of calculated $1^{-}$ resonances is
quoted without details.

Clearly reliable three-body continuum calculations are desirable for
$^{11}$Li, but this is not easy due to both technical difficulties and
lack of information of $^{10}$Li. However, recently a method dealing
with the necessary large distances was formulated for s-waves
\cite{fed93,jen97} and developed and applied for three
$\alpha$-particles and two neutrons surrounding a spin zero core
\cite{fed96}. The purpose of the present letter is to extend the method
to finite core-spin and provide continuum structure calculations for
$^{11}$Li.

\paragraph*{Method.}  The method is described for bound states in
\cite{fed95,fed93,jen97} and for continuum states in \cite{fed96}.  We
shall here briefly sketch the necessary generalizations. The k'th
particle has mass $m_{k}$, charge $eZ_k$, spin ${\bf s}_k$ and
coordinate ${\bf r}_k$. The two-body potentials are $V_{ij}$. The spins
${\bf s}_j$ and ${\bf s}_k$ are coupled to ${\bf s}$ which in turn
coupled to ${\bf s}_i$ gives the total spin. The corresponding wave
function is $\chi_{s}^{(i)}$.  The total wavefunction is expanded in a
complete set of hyperangular functions
\begin{eqnarray} \label{e20}
 \Psi (\rho ,\Omega )=\frac{1}{\rho ^{5/2}}\sum_{n=1}^{\infty}
 f_{n}(\rho ) \frac{\phi _{n}^{(i)}(\rho,\Omega_i)}
{\sin(2\alpha_i)}  \; ,
\end{eqnarray}
where each of the three components $\phi _{n}^{(i)}$ is expressed in
the corresponding set of hyperspherical coordinates ($\rho$,
$\Omega_{i}$) = ($\rho$, $\alpha_i$, $\Omega_{xi}$, $\Omega_{yi}$).
They satisfy the angular part of the Faddeev equations with the
eigenvalues $\lambda_n(\rho)$.

For Borromean systems we select those solutions $\Psi_{n'}$ to
eq.(\ref{e20}) where the large-distance ($\rho \rightarrow \infty)$
boundary conditions for $f^{(n')}_{n}$ are given by \cite{tay72}
\begin{equation} \label{e35}
f^{(n')}_n(\rho) \rightarrow \delta_{n,n'} 
 F^{(-)}_{n}(\kappa \rho) - S_{n,n'} F^{(+)}_{n}(\kappa \rho) \; ,
\end{equation}
where $\kappa^2=2mE/\hbar^2$ and $F^{(\pm)}_{n}$ are related to the
Hankel functions of integer order
\begin{eqnarray}\label{e40}
F^{(\pm)}_{n}(\kappa \rho) = 
 \sqrt{\frac{m \rho}{4 \hbar^2}}\, H^{(\pm)}_{K_n+2}(\kappa \rho)  \; ,
\end{eqnarray} 
where $K_n(K_n+4)=\lambda_n(\rho = \infty)$. The continuum wave
functions are orthogonal and normalized to delta functions in energy.

By diagonalization of the $S$-matrix we obtain eigenfunctions and
eigenphases. Resonance energies and widths can be obtained by use of
the complex energy method, where the Faddeev equations are solved for
$E=E_r-i\Gamma/2$ with the boundary condition $f^{(n')}_n \propto
\delta_{n,n'} \sqrt{\frac{m \rho}{4 \hbar^2}}\,
H^{(+)}_{K_n+2}(\kappa \rho)$ These solutions correspond to poles of
the $S$-matrix \cite{tay72}.

\paragraph*{Large-distance behavior.} We shall now specialize to the 
system of interest, i.e. two neutrons ($k=2,3$) and a core ($k=3$)
with spin $s_c$.  The three independent orbital s-wave components,
$\phi_{L,s_i}^{(i)}, i=1,2,3,$ are characterized by $l_{xi}=0$,
$l_{yi}=L$ and the spin wave functions $\chi_{s_1=0}^{(1)}$,
$\chi_{s_2=s_c-1/2}^{(2)}$, $\chi_{s_3=s_c+1/2}^{(3)}$. All other
partial waves decouple for large $\rho$ over a distance scale defined
by the range of the interactions.  The s-waves, however, are coupled
and feel the interactions over a distance defined by the scattering
lengths.  The remaining two s-wave components are determined by
antisymmetry, i.e. $\phi_{L,s_3}^{(2)} = - \phi_{L,s_3}^{(3)}$,
$\phi_{L,s_2}^{(3)} = \phi_{L,s_2}^{(2)}$. The last combination,
$\phi_{L,s=1}^{(1)}$, is not allowed due to the Pauli principle.

The three components, $\phi_{L,s_i}^{(i)}$ obey for large $\rho$ the
coupled angular Faddeev equations which, to leading order in $\alpha$,
explicitly are given by \cite{fed95,jen97,fed96}
\begin{eqnarray} \label{e50}
  \left(
  \frac{\partial^2}{\partial \alpha^2} + \kappa_i^2(\alpha)
   \right)  \phi_{L,s_i}^{(i)}(\rho,\alpha) =  
   2 \alpha (-1)^L  C^{(i)}_{L} \rho^2 v_{i}(\rho \sin{\alpha}) \; , \\
  \kappa_i^2(\alpha) \equiv 
 -[ \frac{L(L+1)}{\cos^2 \alpha} - \rho^2 v_{i}(\rho \sin{\alpha})
  - \nu^2_n ]  \; ,
\end{eqnarray}
where we defined $\nu^2_ n= \lambda_n +4$, the s-wave interactions
$v_{i}(x_i) = V_{jk}(x_i/\mu_{jk}){2m\over\hbar^2}$,
$m\mu^2_{jk}=m_jm_k/(m_j+m_k)$ and the coupling terms
\begin{eqnarray} \label{e54}
 C^{(i)}_{L} = \sum_{j \neq i} \sum_{s}  C_{s_is}^{ij}  
 \frac{\phi^{(j)}_{L,s}(\rho,\varphi_{k})} {\sin(2 \varphi_{k})} \; ,\;
 C_{ss^\prime}^{ij} = \langle \chi_{s}^{(i)} |
 \chi_{s^\prime}^{(j)} \rangle \; ,
\end{eqnarray}
where $\varphi_k = \arctan \left(m_k(m_i+m_j+m_k)/m_im_j\right)$.  The
$s$-summation runs over the two possible intermediate spin couplings
for the component $j$. The spin overlap matrix elements are diagonal
for $i=j$, i.e.\ $C_{ss^\prime}^{ii} = \delta_{ss^\prime}$ and
symmetric, i.e.\ $C_{ss^\prime}^{ij} = C_{s^\prime s}^{ji}$.

The rescaled potentials $\rho^2 v_i(\rho \sin{\alpha_i})$ approach for
sufficiently large $\rho$ the zero-range potentials, where the
sensitivity to the shape disappears.  Any potential with the same
scattering length and effective range would then lead to results
accurate to the order $\rho^{-2}$. We shall therefore for convenience
use square well potentials \cite{jen97}, where $v_i(\rho
\sin{\alpha_i}) = v_0^{(i)}$ (region I) when $\alpha_i <
\alpha^{(i)}_0 = \arcsin(R_i\mu_{jk}/\rho) \ll 1$ and zero otherwise
(region II).

The solutions to eq.(\ref{e50}) are then for $\alpha_i >
\alpha^{(i)}_0$ given by
\begin{eqnarray}\label{e76}
 \phi_{L,s_i}^{(i,II)}(\rho,\alpha_i) = A_L^{(i)} 
 P_{L}(\nu,\alpha_i) \; , \label{e82}  \\
 P_{L}(\nu,\alpha) \equiv \cos^L\alpha
 \left(
 \frac{\partial}{\partial \alpha} \frac{1}{\cos\alpha}
\right)^L \sin\left[ \nu 
\left(\alpha - \frac{\pi}{2} \right) \right] 
\end{eqnarray}
and for $\alpha_i < \alpha^{(i)}_0$ to leading order in $\alpha_i$
given by
\begin{equation} \label{e89}
\phi_{L,s_i}^{(i,I)}(\rho,\alpha) = B_L^{(i)}\sin(\kappa_i(0)\alpha) 
- 2 \alpha (-1)^L \frac{\rho^2 v_0^{(i)}}{\kappa_i^2(0)}  C^{(i)}_{L} \; ,
\end{equation}
where the wave functions in $C^{(i)}_{L}$ in eq.(\ref{e54}) must be
$\phi^{(i,II)}_{L,s_i}$ 

Matching the solutions, eqs.(\ref{e76}) and (\ref{e89}), and their
derivatives at $\alpha_i=\alpha^{(i)}_0$ gives a linear set of
equations for $A_L^{(i)}$ and $B_L^{(i)}$.  Large-distance physical
solutions are obtained when the corresponding determinant is zero.
This is the quantization condition for $\lambda$ and the eigenvalue
equation determining the asymptotic behavior of $\lambda(\rho)$.

\paragraph*{Input parameters.}  
We use the neutron-neutron interaction which reproduces the low-energy
properties of the nucleon-nucleon system \cite{gar96}. The
neutron-core, n$-^{9}$Li, effective interaction assumes that the spin
of both $^{9}$Li and $^{11}$Li is $\frac{3}{2}$.  We use the mean-field
spin-orbit term ${\bf l}\cdot{\bf s}_n$ and we include a spin-spin term
to differentiate between the two spin couplings for a given orbital
angular momentum $l$. Such a spin splitting term is hard to avoid due
to the strong spin dependence of the underlying basic interaction.
Several phase equivalent parametrizations are possible. They differ in
the number of two-body bound states occupied by the core neutrons which
therefore subsequently must be excluded in the computation
\cite{gar96}. The results are very close and we shall therefore here
only use the shallow potentials without bound states.

We use the radial form, $\exp(-(r/2.55{\rm fm})^2)$, for all
neutron-core potentials. The strengths $V_{s}(l{j_n})$ depend on
orbital $l$ and total neutron angular momentum $j_n$ for zero core
spin, and for finite core spin also on the total spin $s=1,2$.  All
possible s- and p-waves are included whereas other waves can be
ignored to the accuracy we need.  We use $V_1(s_{1/2})=-6.89$ MeV,
$V_2(s_{1/2})=-7.51$ MeV, $V_1(p_{1/2})=-38.59$ MeV,
$V_2(p_{1/2})=-35.65$ MeV, $V_1(p_{3/2})=43.91$ MeV and
$V_2(p_{3/2})=46.85$ MeV corresponding to virtual s-states at 0.247
MeV and 0.140 MeV and $p_{1/2}$-resonances at 0.75 MeV and 1.60 MeV.
The $p_{3/2}$-potential is purely repulsive thereby simulating the
Pauli blocking by the core neutrons.

The choice of these parameters is dictated by the accumulated
information about the structure of $^{10}$Li, i.e. a p-resonance at
about 0.6 MeV, a low-lying virtual s-state and a small spin splitting
of these states \cite{zin97,abr95}.  We also demand that the
$^{11}$Li-binding energy is reproduced; we obtain 305 keV with the
corresponding root mean square radius of 3.34 fm.  The calculated
fragment momentum distributions in break-up reactions also compare
rather well with measured values suggesting about 20\% of the
$p^2$-configuration in the $^{11}$Li ground state wave function
\cite{gar96}. The remaining part is in the
$s^2$-configuration. Limited ambiguity, although some, is then left
for the parameters of these ``realistic'' interactions.

The number of Jacobi polynomials in the basis expansion is carefully
chosen to give accurate numerical results up to a distance, typically
around 40 fm, where the asymptotic behavior is reached and from then
on the asymptotic solutions eqs.(\ref{e76})-(\ref{e89}) are
used. The accurate low-energy continuum spectrum calculations require
integration of the radial equations up to distances of the order of
five times the sum of the scattering lengths. For the n+n+$^{9}$Li
system this is about 200 fm. 

\begin{figure}[ht]
\centerline{
\psfig{figure=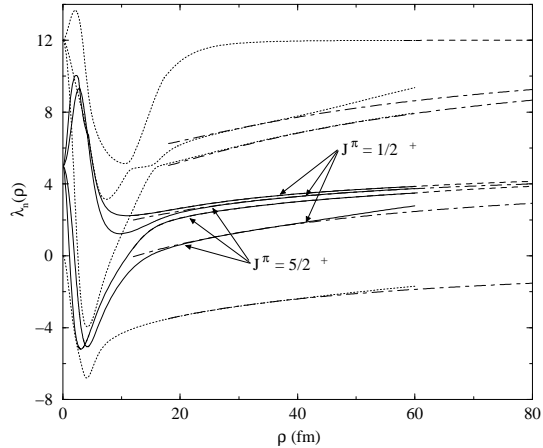,bbllx=25mm,bblly=120mm,bburx=110mm,bbury=190mm,
clip=,width=75mm}}
\caption{The lowest angular eigenvalues $\lambda_n$ as functions of
$\rho$ for angular momentum $J^{\pi}=\frac{3}{2}^{-}$ (dotted lines),
and $\frac{3}{2}^{+}$ (solid lines) for $^{11}$Li. The dot-dashed lines
are the large distance asymptotic behavior. The levels marked
$\frac{1}{2}^{+}$ and $\frac{5}{2}^{+}$ also appear for those quantum
numbers.  The neutron-neutron interaction is from \protect\cite{gar96}
and the neutron-$^9$Li interaction is given in the text. Maximum
hyperspherical quantum numbers up to $K_n \approx 100$ are used.}
\end{figure}

\paragraph*{Angular eigenvalues.}
The key quantities are the angular eigenvalues $\lambda_n(\rho)$,
which, divided by $\rho^2$, are the adiabatic effective radial
potentials. They depend on total angular momentum and parity as seen
in Fig. 1, where we show the spectra for $\frac{3}{2}^{-}$ (ground
state) and $\frac{3}{2}^{+}$ ($1^{-}$-excitation). The structure is
complicated at small distances where avoided level crossings are
seen. The lowest level has in both cases an attractive pocket, which
is responsible for the ground state ($\frac{3}{2}^{-}$) and several
resonances. At large distance the structure is simpler as the
hyperspherical spectrum is approached. In the computation we use the
asymptotic behavior (shown in Fig. 1) of the three coupled s-waves,
one from each Jacobi set.

The $^{11}$Li-states can be labeled as arising from couplings of the
$s_{1/2}$ and $p_{1/2}$ neutron-core states to angular momentum and
parity $0^{\pm}$ and $1^{\pm}$, which in turn is coupled with the core
spin $3/2^{-}$. The result for each parity $\pi=\pm 1$ is three nearly
degenerate states of $1/2^{\pi}$, $3/2^{\pi}$, $5/2^{\pi}$ and one
non-degenerate $3/2^{\pi}$-state.  This structure is reflected in the
eigenvalue spectra where $3/2^{\pm}$ contain additional levels on top
of spectra similar to those of $1/2^{\pm}$ and $5/2^{\pm}$.  The
degeneracy would be complete without spin splitting of the
neutron-core potentials.

\begin{figure}[t]
\centerline{
\psfig{figure=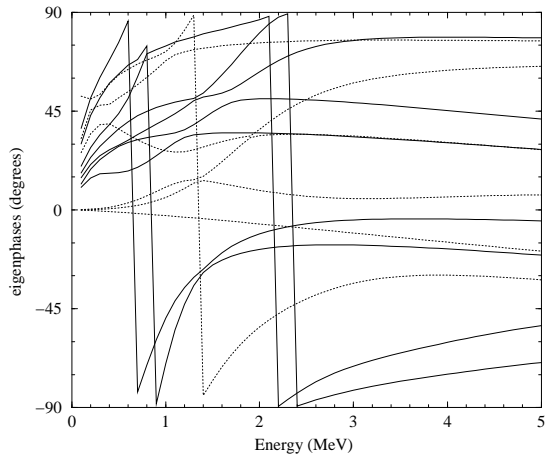,bbllx=25mm,bblly=120mm,bburx=110mm,bbury=190mm,
clip=,width=75mm}}
\caption{The eigenphases corresponding to the lowest $\lambda_n$-values
obtained after diagonalization of the $S$-matrix for
$J^{\pi}=\frac{3}{2}^{-}$ (dotted  lines) and $\frac{3}{2}^{+}$ (solid
lines). The interactions are the same as in Fig. 1.}
\end{figure}

\paragraph*{Phase shifts, poles and the Efimov effect.} 
We show in Fig. 2 the eigenphases for the cases in Fig. 1. For
$\frac{3}{2}^{-}$ we find rapid variations at smaller energy and one
crossing of $\pi/2$ at about 1.2 MeV. For $\frac{1}{2}^{-}$ and
$\frac{5}{2}^{-}$, none of them shown, we find rather smooth phase
shifts indicating that the structure in $\frac{3}{2}^{-}$ is due to the
non-degenerate components.  For $\frac{3}{2}^{+}$ four crossings of
$\pi/2$ are seen in two pairs.  In $\frac{1}{2}^{+}$ and
$\frac{5}{2}^{+}$ is found a similar structure, but one member of each
pair is not present indicating the non-degenerate nature.

\begin{table}[ht]
\renewcommand{\baselinestretch}{0.9}
\caption{The real and imaginary values $(E_r,\Gamma)$ (in MeV) of the
lowest $S$-matrix poles $E=E_r-i\Gamma/2$ for $^{11}$Li for various
spins and parities $J^{\pi}$. The excitation energy $E^*=E_r + 0.305$
MeV.  The interactions for the upper part of the table are the same as
in Fig. 1. The lower part of the table are results for a model with
$s_c=0$ and the same average positions of the neutron-core
resonances. i.e. energies of the $s_{1/2}$ virtual state and the
$p_{1/2}$ resonance at 0.2 MeV and 1.22 MeV, respectively.  }
\renewcommand{\baselinestretch}{1.5} 
\begin{center}
\begin{tabular}{c|cc|cc|cc|cc|cc|}
$J^{\pi}$ & $E_r$ & $\Gamma$ & $E_r$ & $\Gamma$ & $E_r$ & $\Gamma$ 
& $E_r$ & $\Gamma$ & $E_r$ & $\Gamma$ \\ 
\hline 
$\frac{1}{2}^{-}$ &-&-&-&-&1.37&0.51&1.56&0.56&1.98 & 0.65 \\
$\frac{3}{2}^{-}$ &-0.305&0&0.89&0.43&1.41&0.56&1.60&0.61&2.03&0.68 \\
$\frac{5}{2}^{-}$ &-&-&-&-& 1.36 & 0.49 & 1.60 & 0.68 & 2.01 & 0.72 \\
\hline 
$\frac{1}{2}^{+}$ &0.65&0.35&-&-&1.28&0.48 & 1.74 & 0.64 & 1.95 & 0.68 \\
$\frac{3}{2}^{+}$ &0.68&0.33&0.88&0.33&1.33&0.50&1.77&0.63&2.08 & 0.71 \\
$\frac{5}{2}^{+}$ &0.68&0.37&-&-&1.36&0.55&1.74& 0.64 & 2.11 & 0.84 \\
\hline 
\hline 
$0^{+}$ & -0.305 &0&1.00&0.37& 1.35 & 0.45 & 1.62 & 0.61 & 1.96 & 0.92 \\
$1^{+}$ & - & - & - & - & 1.40&0.59 & 1.59 & 0.63 & 2.02 & 0.81 \\
\hline 
$0^{-}$ & - & - & 0.92 & 0.39 & 1.25 & 0.51 & 1.82 & 0.62 & 2.02 & 0.65 \\
$1^{-}$ & 0.64 & 0.31 & - & - & 1.46 & 0.53 & 1.76 & 0.59 & 2.08 & 0.67 \\
\end{tabular} 
\end{center}
\label{tab1} 
\end{table}

For the lowest spins and parities we give in Table 1 the lowest
S-matrix poles obtained by the complex energy method.  We see the three
times nearly degenerate $1^{-}$-resonance at about 0.65 MeV with a
width of about 0.35 MeV. We also see degenerate $0^{\pm}$-resonances at
0.89 MeV with widths of 0.33 MeV and 0.43 MeV. In addition degenerate
$1^{\pm}$-poles appear at 1.3 - 1.4 MeV, 1.6 - 1.8 MeV and 1.9 - 2.1
MeV with widths of about 0.5 MeV, 0.6 MeV and 0.8 MeV, respectively.
Also poles of $0^{\pm}$-type should be present for $3/2^{\mp}$ at these
energies as seen by comparing with the model of zero core spin. They
can not be distinguished numerically.

The height of the effective radial barriers are 1.7 MeV, 0.9 MeV, 0.7
MeV and 0.6 MeV for $\frac{1}{2}^{-}$, $\frac{1}{2}^{+}$,
$\frac{3}{2}^{+}$ $\frac{5}{2}^{+}$, respectively. The pocket and the
barrier in this potential is absent for both $\frac{1}{2}^{-}$ and
$\frac{5}{2}^{-}$. These potentials are all attractive when the
centrifugal barrier is removed.  The lowest resonances appear around
the barrier and their widths are consequently relatively large and
rather sensitive to fine tuning of the interactions. Related cross
sections would probably be relatively smooth.  We have not attempted
to reproduce a $1^{-}$-resonance precisely at 1 MeV \cite{kor97}, but
we still reproduce the properties almost within the experimental
uncertainties.

The relatively large number of additional poles could be due to the
Efimov effect, which is an anomaly in a three body system when the
scattering lengths are much larger that the range of the interactions
\cite{efi70,fed93}.  With increasing scattering lengths, the
infinitely many poles of the three-body S-matrix move towards the
point $E$=0. For very large but finite scattering lengths a number of
poles must already appear close to zero.  These poles originate from
the long distance tail of the effective potential ($ \propto
\sum_{i=1,3}{a_i}\mu_{jk}^{-1}\rho^{-3}$, where $a_i$ is the
scattering length of the $i$-th subsystem) and they are not sensitive
to the details of the interactions.  Since there are no confining
barriers for these poles, their corresponding widths must be rather
large.

In our case the Efimov condition is almost fulfilled, since the
scattering lengths are much larger then the range of the interactions:
$a_{nn}\mu_{nn}^{-1}+2a_{cn}\mu_{cn}^{-1}\approx$ 50~fm.  This must
necessarily result in a number of broad resonances near the $E$=0
point.

\paragraph*{Transition strengths and Coulomb cross sections.}
Dipole strength functions for different potentials are shown in Fig. 3
both for computed continuum wave functions and for plane waves.  The
interaction without spin splitting (approximation with zero core spin)
gives a distribution shifted about 100 keV towards lower energy
compared to the result for the realistic full computation. A lower and
broader peak is obtained for the potential from \cite{joh90} where the
$p^2$-content of the three-body wave function is very small. For
comparison we also show the result for the same potential with the
limited basis employed in \cite{dan94}. The low-lying
$1^{-}$-resonances enhance the strength functions at low energies
compared to the plane wave computation. The computed strength
functions substantially exceed most of the data points
\cite{zin97,sac93,shi95} in the peak region around 0.55 MeV. (Note
that the data in \cite{zin97,sac93} contain much less total strength.)
A reduction could be achieved with higher resonance energy and larger
width, but this would probably only be provided by a potential with
much too small $p^2$-content in the three-body wave function. On the
other hand, the proper comparison with the different experimental
results is not obvious.

\begin{figure}[t]
\centerline{
\psfig{figure=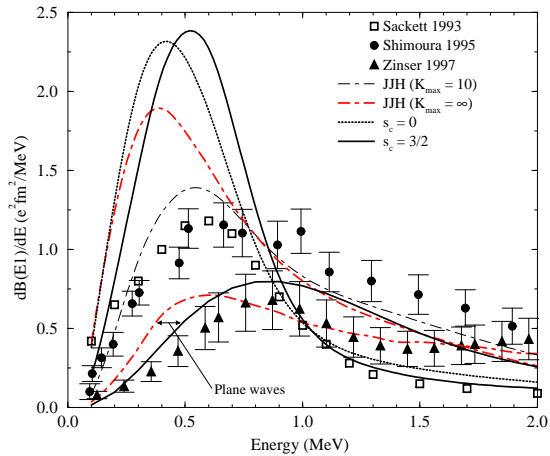,bbllx=25mm,bblly=120mm,bburx=110mm,bbury=190mm,
clip=,width=75mm}}
\caption{The dipole strength functions $dB/dE$ are shown
here for several potentials for comparison.  The curves peaked at low
energy are for the correct continuum wave functions and the broader
curves are for plane waves.  The interactions for $s_c=3/2$ are the
same as in Fig. 1 and for $s_c=0$ as described in table 1.  JJH refers
to the neutron-core potential from \protect\cite{joh90}, where the
result for the limited basis $K_{max}=10$ is from
\protect\cite{dan94}.  The measured curves are from
\protect\cite{zin97,sac93,shi95}.  We normalize the data in
\protect\cite{shi95} to our sum rule value, while the absolute data
from \protect\cite{zin97,sac93} is left unchanged.}
\end{figure}

The Coulomb dissociation cross section is obtained by folding the
virtual photon spectrum with the strength function. The precise
expressions for the dipole approximation adopted here can be found in
\cite{ber88}. The low-energy enhancement places the strength where the
virtual photon spectrum is relatively large and therefore necessarily
also implies a large Coulomb cross section, since the dipole is the
dominating dissociation mode. The total cross sections for beam
energies 42 MeV/A, 180 MeV/A, 280 MeV/A and 800 MeV/A on a lead target
are 4.487 b, 2.128 b, 1.429 b and 0.971 b, respectively. This is in
agreement with the available measured values \cite{zin97}, but
substantially larger than 2.586 b, 1.166 b, 0.933 b and 0.657 b found
by using plane waves. For 280 MeV/A we find 1458 b, 1501 b, 1283 b,
respectively for the potentials without spin splitting, with small
$p^2$-content with complete and  limited basis.

\paragraph*{Conclusions.} A recently formulated method was used to
compute low-energy three-body continuum spectra for $^{11}$Li for
various angular momentum states.  The angular part of the Faddeev
equations are treated numerically at short distances, whereas the
large distance behavior of eigenvalues and eigenfunctions is computed
essentially analytically. Combining the results from these two regions
allow accurate computations at large distances.

We employ interactions that reproduce reasonably well the low energy
continuum properties of the two-body subsystems and the fragment
momentum distributions in break-up reactions.  The correct spins of
$^{9}$Li and $^{11}$Li are used implying level splitting and fine
structure beyond the approximation of zero core spin. A
$1^{-}$-resonance at about 0.65 MeV with a width of 0.3 MeV is found
in fair agreement with a recent measurement.  Also several other
s-matrix poles were found and related to the Efimov effect. Strength
functions, Coulomb dissociation cross sections, phase shifts and
S-matrix poles for $J^{\pi}=\frac{1}{2}^{\pm}, \frac{3}{2}^{\pm},
\frac{5}{2}^{\pm}$ are computed. A proper comparison of measured and
computed strength functions would be rewarding either by showing the
limitations of, or by selecting, the most correct three-body model.

\paragraph*{\bf Acknowledgments.} A.C. acknowledges support by
the European Union via the Human Capital and Mobility program
contract nr. ERBCHBGCT930320.

\end{document}